\documentclass[aps,prd,twocolumn,groupedaddress]{revtex4-1}
\usepackage{graphicx}
\include{amssym}

\begin{document}
\title{Anomalous decay $f_1(1285)\to\pi^+\pi^-\gamma$ in the Nambu - Jona-Lasinio model}
\author{A. A. Osipov}
\email[]{aaosipov@theor.jinr.ru} 
\altaffiliation{}
\affiliation{Bogoliubov Laboratory of Theoretical Physics, Joint Institute for Nuclear Research, Dubna, 141980, Russia}

\author{M. K. Volkov}
\email[]{volkov@theor.jinr.ru} 
\thanks{}
\affiliation{Bogoliubov Laboratory of Theoretical Physics, Joint Institute for Nuclear Research, Dubna, 141980, Russia}

\begin{abstract}
Using the Nambu--Jona-Lasinio model with the $U(2)\times U(2)$ chiral symmetric effective four-quark interactions, we derive the amplitude of the radiative decay $f_1(1285) \to\pi^+\pi^-\gamma$, find the decay width $\Gamma (f_1\to\pi^+\pi^-\gamma)=346\,\mbox{keV}$ and obtain the spectral di-pion effective mass distribution. It is shown that  in contrast to the majority of theoretical estimates (which consider the $a_1(1260)$ meson exchange as the dominant one), the most relevant contribution to this process comes out from the $\rho^0$-resonance exchange related with the triangle $f_1\rho^0\gamma$ anomaly. The spectral function is obtained to be confronted with the future empirical data.     
\end{abstract}

\maketitle

\section{Introduction}
Recently \cite{Osipov17prd} the anomalous radiative decay $f_1\to\rho^0\gamma$ of the axial-vector $I^G(J^{PC})=0^+(1^{++}$) $f_1(1285)$ meson has been described on the basis of the effective Lagrangian obtained as a result of the derivative expansion of the underlying triangle quark diagram. The Nambu--Jona-Lasinio (NJL) model with $U(2)\times U(2)$ chiral symmetry spontaneously broken down to the diagonal $SU(2)_I \times U(1)_V$ subgroup \cite{Volkov84,Volkov86,Ebert86,Bijnens93,Bernard96,Osipov17ap} has been used (the quantum anomaly breaks the axial $U(1)_A$ symmetry). The model not only reproduces the nontrivial structure of the $f_1\rho\gamma$ vertex (which accumulates the most general restrictions imposed on it by gauge symmetry and statistics \cite{Rosenberg63,Adler69}), but also fixes fully the values of the coupling constants involved. The decay width found, $\Gamma_{f_1\to\rho\gamma}=311\ \mbox{keV}$, is compatible with the recently published value $\Gamma_{f_1\to\rho^0\gamma}=453\pm 177\ \mbox{keV}$ \cite{Dickson16} measured for the first time in photoproduction from a proton target using the CLAS detector at Jefferson Laboratory, and four times less the value $\Gamma_{f_1\to\rho^0\gamma}=1326\pm 313\ \mbox{keV}$, quoted by the Particle Data Group \cite{Patrignani16}. The $1/N_c$ arguments have been used to explain why the derivative expansion is suitable for the phenomenological description of the $f_1\to\rho^0\gamma$ decay amplitude.    

Here we apply the same approach to describe the radiative decay $f_1(1285)\to\pi^+\pi^-\gamma$. Apart from the $AVV$ triangle anomaly, which has been newly analysed by the CLAS Collaboration \cite{Dickson16}, this process contains information on the triangle $AAA$ and box $AAAV$ anomalies. It would be interesting to subject both anomalies to experimental scrutiny. Unfortunately, the decay width of the $f_1(1285)\to\pi^+\pi^-\gamma$ transition is not measured yet, although a clear signal of $f_1(1285)$ has been seen in the effective mass spectrum of the $\pi^+\pi^-\gamma$ system in the reaction $\pi^- N\to\pi^-\pi^+\pi^-\gamma N$ at a pion beam with the momentum $p_{\pi^-}=37\,\mbox{GeV/c}$ studied at the VES spectrometer of IHEP \cite{Amelin95}. The centrally produced exclusive final states formed in the reaction $pp\to p_f(\pi^+\pi^-\gamma )p_s$ at 300 GeV/c have been studied by the WA76 Collaboration at CERN Omega Spectrometer \cite{Armstrong92}. The $\pi^+\pi^-\gamma$ mass spectrum shows two enhancements, one at 0.96 GeV due to the $\eta'(958)$ and one at 1.27 GeV which could be due to the $f_1(1285)$. This WA76 Collaboration, in particular, has measured the $\pi^+\pi^-$ mass spectrum from the reaction $f_1(1285)\to\pi^+\pi^-\gamma$, where a $\rho^0 (770)$ signal can clearly be seen. Events $\gamma p\to pf_1(1285)\to p\pi^+\pi^-\gamma$ were also identified by CLAS \cite{Dickson16} using kinematic fitting and time-of-flight selections.  
   
An earlier attempt of the theoretical description of the $f_1(1285)\to\pi^+\pi^-\gamma$ decay width can be found in \cite{Li95}. It is made in the framework of the $U(2)_L\times U(2)_R$ chiral theory of mesons. In this approach the amplitude gets two types of contributions: the direct coupling $f_1\to\pi\pi\gamma$ and the $a_1(1260)$-meson exchange $f_1\to a_1\pi\to\pi\pi\gamma$. The  $\rho$ exchange was not taken into account because the Bardeen's form of the non-abelian anomaly \cite{Schechter84a,Schechter84b,Schechter85} used there forbids the $f_1\to\rho^0\gamma$ transition. As a result, an estimate $\Gamma_{f_1\to\pi\pi\gamma}=18.5\ \mbox{keV}$ has been obtained. A consistent scheme \cite{Kugo85,Meissner90} contains the anomalous $f_1\rho^0\gamma$ vertex. Its contribution through the channel $f_1\to\rho^0\gamma\to\pi\pi\gamma$ can be more important than the $a_1$ exchange. The reason for this is very simple. The $a_1$ exchange is described by two propagators $(m_{a_1}^2-t)^{-1}$, or $(m_{a_1}^2-u)^{-1}$. The physical region of the kinematical variables $t$ and $u$ is given by $m_\pi\le \sqrt{t}, \sqrt{u}\le (m_{f_1}-m_\pi )$. It is clear that both propagators have no poles at physical values of meson masses. Contrary to this, the propagator of the $\rho$ meson contributes as $(m_\rho^2-s)^{-1}$, where $2m_\pi \le \sqrt{s}\le m_{f_1}$. Thus, it has a pole on the real axis (to lowest order in $1/N_c$, where $N_c$ is a number of colors). This certainly indicates that only a contribution from the $a_1$ exchange is not sufficient.

The purpose of this paper is to examine the role of the vector $\rho(770)$ and axial-vector $a_1(1260)$ resonances in the radiative decay $f_1(1285)\to\pi^+\pi^-\gamma$ in more detail. On one hand, our treatment of the problem is beyond the simplified version employed in \cite{Li95}, on the other hand, our calculations of the di-pion mass spectrum, can be further improved by taking into account the non-perturbative effects of final-state interactions. However, due to the absence of empirical data we postpone the calculations of these $1/N_c$ suppressed contributions. 

The paper is organized as follows. In Section \ref{ELag} we discuss the quark-meson Lagrangian of the NJL model and fix the coupling constants.  In Section \ref{Ampl} we obtain the $f_1(1285)\to\pi^+\pi^-\gamma$ decay amplitude. Our estimates of the decay width and di-pion spectral function are presented in Section \ref{DecW}. Finally, we summarize and make conclusions in Section \ref{concl}. 

\section{Effective Lagrangian}
\label{ELag}
The main tool of our study is the bosonized version of the NJL model with $U(2)\times U(2)$ chiral symmetric four-quark interactions. The effective meson vertices can be obtained through the derivative expansion of the underlying one-quark-loop diagrams. This can be done both in momentum space \cite{Volkov84,Volkov86,Bernard96} or in position space \cite{Ebert86,Bijnens93,Osipov17ap}. In the latter case, the heat kernel technique adjusts the derivative expansion of the quark determinant in such a way that chiral symmetry is protected for each Seeley-DeWitt coefficient. The result is well known, thus we will write down here only the part which is responsible for quark-meson interactions governing the $f_1(1285)\to\pi^+\pi^-\gamma$ decay. The corresponding Lagrangian density is
\begin{eqnarray}
\label{Lint}
{\cal L}_{int} &=& \bar q\left\{ i\gamma^\mu(\partial_\mu -ieA_\mu )- M +ig_\pi\gamma_5 \vec\pi\vec\tau \right. \nonumber \\
&+& \left.\frac{g_\rho}{2}\gamma^\mu  \left[\vec\rho_\mu\vec\tau +\gamma_5(\vec a'_{1\mu}\vec\tau+f_{1\mu}\tau_0)\right]\right\}q.  
\end{eqnarray}
Our notations are as follows: $\vec\tau$ are the standard $SU(2)$ Pauli matrices, $\tau_0$ is a unit $2 \times 2$ matrix in the isospin space;  $\gamma^\mu$ and $\gamma_5$ are Dirac matrices in a four dimensional Minkowski space. The light quark fields $q=(u,d)$ have color and 4-spinor indices which are suppressed. The diagonal matrix $M=m\tau_0$, where $m=276\,\mbox{MeV}$ is a constituent quark mass, preserves isospin symmetry, i.e. $m=m_u=m_d$. $A_\mu =Q{\cal A}_\mu$, where ${\cal A}_\mu$ is the electromagnetic field, and $Q$ is the matrix of the light quark's charges in relative units of the proton charge $e$ 
\begin{equation}
Q=\frac{1}{2}\left(\tau_3 +\frac{1}{3}\right).
\end{equation}
The $\vec\pi$ and $\vec\rho_\mu$ are the field operators associated with the iso-triplet of pions $\pi(140)$ and vector $\rho (770)$-mesons; $f_{1\mu}$ describes the iso-singlet axial-vector $f_1(1285)$-meson, and $\vec a'_{1\mu}$ stands for the unphysical axial-vector field that should be redefined to avoid the $\vec\pi -\vec a'_{1\mu}$ mixing
\begin{equation}
\label{pa-trans}
\vec a'_{1\mu}=\vec a_{1\mu}+\sqrt{\frac{2Z}{3}}\kappa m\partial_\mu\vec\pi.
\end{equation}   
Here $Z=(1-6m^2/m_{a_1}^2)^{-1}=1.4$, and a dimensionful parameter $\kappa$ is fixed by requiring that the Lagrangian does not contain the $\vec\pi-\vec a_{1\mu}$ transitions, it gives $\kappa =3/m_{a_1}^2$. The $\vec a_{1\mu}$ field represents a physical axial-vector state $a_1(1260)$ of mass $m_{a_1}=1230\pm 40\,\mbox{MeV}$. In our estimates we take the value $m_{a_1}=1264\,\mbox{MeV}$. 

Since the free part of the meson Lagrangian following from evaluation of the one-quark-loop self-energy diagrams must preserve its canonical form, one should renormalize the bare meson fields by introducing the Yukawa coupling constants $g_\pi$ and $g_\rho$ in eq. (\ref{Lint}). To absorb infinities of self-energy graphs, these couplings depend on the divergent integral which is regularized in a standard way \cite{Osipov17ap}
\begin{equation}
\label{j1}
J_1(m^2)=\ln\left(1+\frac{\Lambda^2}{m^2}\right)-\frac{\Lambda^2}{\Lambda^2+m^2}.   
\end{equation}
A finite ultraviolet cutoff $\Lambda =1250\,\mbox{MeV}$ restricts the region of integration in the quark-loop integrals and characterizes the energy scale where the NJL model is applicable. Thus, we have   
\begin{equation}
\label{g-pi-rho}
g_\pi=\sqrt{Z}g,\quad  g_\rho=\sqrt{6}g, \quad g^2=\frac{4\pi^2}{N_cJ_1}.
\end{equation}
The coupling $g_\pi$ satisfies the quark analog of the Goldberger -- Treiman relation, which is $m=f_\pi g_\pi$, where $f_\pi=93\,\mbox{MeV}$ is the weak pion decay constant.  $g_\rho$ is the coupling of the $\rho\to\pi\pi$ decay ($g_\rho^2/4\pi =\alpha_\rho =3$). 

The two non-anomalous vertices needed for our calculations are \cite{Osipov17ap}
\begin{eqnarray}
\label{rpp}
{\cal L}_{\rho\pi\pi}&=&-i\frac{g_\rho}{4}\,\mbox{tr}\left(\rho_\mu [\pi , \partial^\mu\pi ] \right.\nonumber\\ 
&-&\left.\frac{Z-1}{2m_{a_1}^2}\,\rho_{\mu\nu}[\partial^\mu\pi , \partial^\nu\pi ]\right), \\
\label{apg}
{\cal L}_{a_1\pi\gamma}&=&\frac{i}{4}f_\pi e g_\rho Z \,\mbox{tr}\left\{a_1^\mu [A_\mu , \pi ] \right. \nonumber \\
&+&\left.\frac{1}{m_{a_1}^2}\left(F_{\mu\nu}[a_1^\mu , \partial^\nu\pi ]+a_1^{\mu\nu}[A_\mu , \partial_\nu\pi]\right)\right\},
\end{eqnarray}   
where $\rho_\mu =\vec\rho_\mu\vec\tau$, $a_{1\mu}=\vec a_{1\mu}\vec\tau$, $\pi =\vec\pi\vec\tau$;  the quantities $\rho_{\mu\nu}, a_{1\mu\nu}, F_{\mu\nu}$ stand for the field strengths $\rho_{\mu\nu}=\partial_\mu\rho_\nu-\partial_\nu\rho_\mu$, $a_{1\mu\nu}=\partial_\mu a_{1\nu}-\partial_\nu a_{1\mu}$, and $F_{\mu\nu}=\partial_\mu A_\nu-\partial_\nu A_\mu$. In the following we neglect the second term in (\ref{rpp}). On the $\rho$-meson mass shell it has a small factor $(Z-1)(m_\rho /m_{a_1})^2/2=0.075$ (compared with the factor 1 of the first term).  
 
Following \cite{Osipov17prd}, we may write down the effective Lagrangian density, which describes the  $f_1\rho^0\gamma$ anomalous transition   
\begin{eqnarray}
\label{lag-f}
{\cal L}_{f_1\rho^0\gamma}&=&-\frac{e\alpha_\rho}{8\pi m^2}e^{\mu\nu\alpha\beta} \left(\rho^0_{\mu\nu}F_{\alpha\sigma}\partial^\sigma f_{1\beta} \right. \nonumber \\
&+&\left.\frac{1}{2}f_{1\mu\nu}F_\alpha^{\ \,\sigma}\rho^0_{\sigma\beta}+F_{\mu\nu}\partial^\sigma \rho^0_{\sigma\alpha}f_{1\beta}\right)   
\end{eqnarray}
with the strength tensor $f_{1\mu\nu}=\partial_\mu f_{1\nu} -\partial_{1\nu} f_\mu$.  

The other anomalous vertex describes the $f_1a_1\pi$ interaction
\begin{equation}
\label{fap}
{\cal L}_{f_1a_1\pi}=g_{f_1a_1\pi} e^{\alpha\beta\mu\nu} f_{1\alpha} \partial_\mu\vec a_{1\beta} \partial_\nu\vec\pi ,
\end{equation} 
where 
\begin{equation}
g_{f_1a_1\pi}=\frac{\alpha_\rho}{2\pi f_\pi}\left[1+(1-3a)\frac{Z-1}{2Z}\right].
\end{equation}
The second term in the brackets is due to the replacement (\ref{pa-trans}). The derivative coupling  $\bar q\gamma^\mu\gamma_5\partial_\mu\pi q$ makes the corresponding triangle quark diagram linearly divergent. A superficial linear divergence appears in the course of evaluation of the overall finite integral. Shifts in the internal momentum variable of the closed fermion loop integrals induce an arbitrary finite  surface term contribution  proportional to $(1-3a)$. Here $a$ is a dimensionless constant, controlling the magnitude of an arbitrary local part \cite{Jackiw00,Hiller01}. Hereinafter it will be fixed by the requirement of gauge invariance of the $f_1(1285)\to\pi^+\pi^-\gamma$ decay amplitude.

\section{Amplitude of the $f_1(1285)\to\pi^+\pi^-\gamma$ decay}
\label{Ampl}
The Lagrangian density (\ref{Lint}) generates three different types of contributions to the amplitude of the reaction $f_{1}(l)\to\pi^+(p_+)\pi^-(p_-)\gamma (p)$. These are the contributions through the intermediate $\rho^0$ and $a_1^\pm$ mesons and a contact interaction induced by the quark box diagram (Figs.1-3). The corresponding amplitude can be written as follows
\begin{eqnarray}
\label{amplitude}
T&=& ie_{\mu\nu\alpha\beta} \epsilon^\beta (l)\epsilon^*_\gamma (p)\left[ g^{\alpha\gamma}\left( F_1l^\mu p^\nu_+ +F_2l^\mu p^\nu_- + F_3 p_+^\mu p_-^\nu \right)\right.  \nonumber\\
&+& \left. F_4 p^\alpha l^\gamma p_+^\mu p_-^\nu \right]
\end{eqnarray}
where $\epsilon_\beta (l)$,  $\epsilon_\gamma (p)$ are the polarization vectors of the $f_1$-meson and the photon; $l$, $p$, $p_+$, $p_-$ are the 4-momenta of $f_1$-meson, photon and charge pions. For the respective form factors $F_a$ $(a=1,2,3,4)$ the diagram with the exchange of the $\rho^0$-meson (Fig.1) gives 
\begin{eqnarray}
F_1^{(\rho )}&=& \frac{3eg_\rho}{Z(4\pi f_\pi )^2}\frac{m_{f_1}^2-2m_{f_1}(\varepsilon -\varepsilon_- )}{m_\rho^2 - s -im_\rho \Gamma_\rho (s)},  \\
F_2^{(\rho )}&=&\frac{-3eg_\rho}{Z(4\pi f_\pi )^2}\frac{m_{f_1}^2-2m_{f_1}(\varepsilon-\varepsilon_+ )}{m_\rho^2 - s -im_\rho \Gamma_\rho (s)}, \\
F_3^{(\rho )}&=&\frac{3eg_\rho}{Z(2\pi f_\pi )^2}\frac{m_{f_1}^2-m_{f_1}\varepsilon}{m_\rho^2 - s -im_\rho \Gamma_\rho (s)}, \\
F_4^{(\rho )}&=&\frac{-3eg_\rho}{2Z(2\pi f_\pi )^2}\frac{1}{m_\rho^2 - s -im_\rho \Gamma_\rho (s)}, 
\end{eqnarray}
where $s=(l-p)^2$, $\varepsilon, \varepsilon_\pm$ are the energies of photon and charged pions in the rest frame of the $f_1$ meson, and $\Gamma_\rho (s)$ is the hadronic off-shell width of the $\rho (770)$ resonance \cite{Pich00}
\begin{equation}
\Gamma_\rho (s)=\frac{m_\rho s}{96\pi f_\pi^2}\left[\sigma_\pi^3\theta (s-4m_\pi^2) +\frac{1}{2}\sigma_K^3 (s-4m_K^2) \right],
\end{equation}
with $\sigma_P=\sqrt{1-4m_P^2/s}$, $m_\pi =138\,\mbox{MeV}$ and $m_K=494\,\mbox{MeV}$. We find that the contribution of this channel is $\Gamma (f_1\to\rho^0\gamma\to\pi^+\pi^-\gamma )=276\ \mbox{keV}$. One can use the sequential decay formula to do a cross check
\begin{equation}
\Gamma (f_1\to\rho^0\gamma\to\pi^+\pi^-\gamma )=\frac{\Gamma (f_1\to \rho^0\gamma )\Gamma (\rho^0\to\pi^+\pi^-)}{\Gamma_\rho} .
\end{equation}
Since $\rho$ decays into $\pi\pi$ to hundred percent, and it is known from our previous estimates that $\Gamma (f_1\to \rho^0\gamma ) = 311\,\mbox{keV}$ \cite{Osipov17prd} we conclude that both results are in good agreement. Notice that for simplicity we could ignore the energy dependence in $\Gamma_\rho (s)$ taking its empirical value $\Gamma_\rho =149.1\pm 0.8\,\mbox{MeV}$. This would diminish the result only on 1\%. 

\begin{figure}[htb]
\label{figura1}
\includegraphics[width=0.45\textwidth]{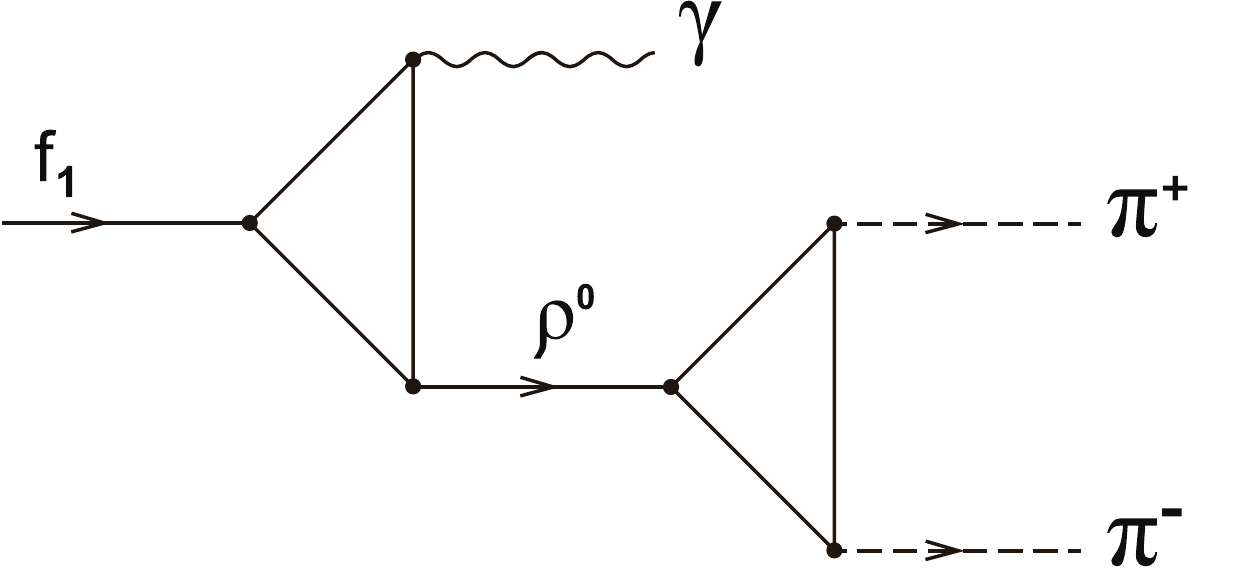}
\caption{The contribution through the intermediate $\rho^0$ meson to the radiative decay amplitude $f_{1}\to \pi^+\pi^-\gamma$. The first triangle diagram is described by the effective Lagrangian density (\ref{lag-f}), the second one by the Lagrangian density (\ref{rpp}).}
\end{figure} 

\begin{figure}[htb]
\label{figura2}
\includegraphics[width=0.45\textwidth]{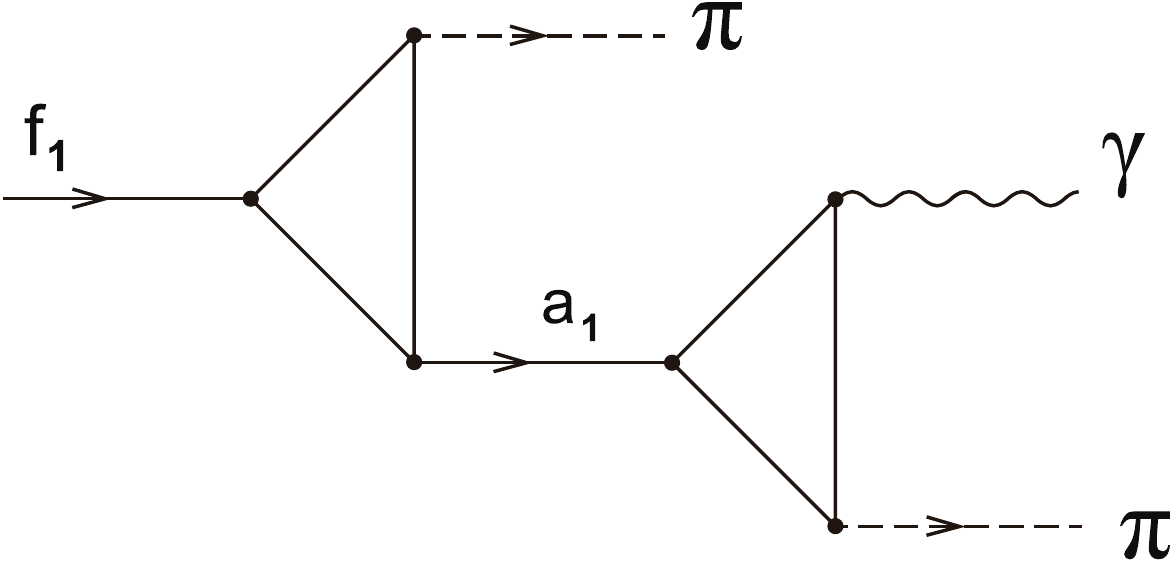}
\caption{The contribution through the intermediate $a_1(1260)$ meson to the radiative decay amplitude $f_{1}\to \pi^+\pi^-\gamma$. The first triangle diagram is described by the effective Lagrangian density (\ref{fap}), the second one by the Lagrangian density (\ref{apg}).}
\end{figure} 

\begin{figure}[t]
\label{figura3}
\includegraphics[width=0.45\textwidth]{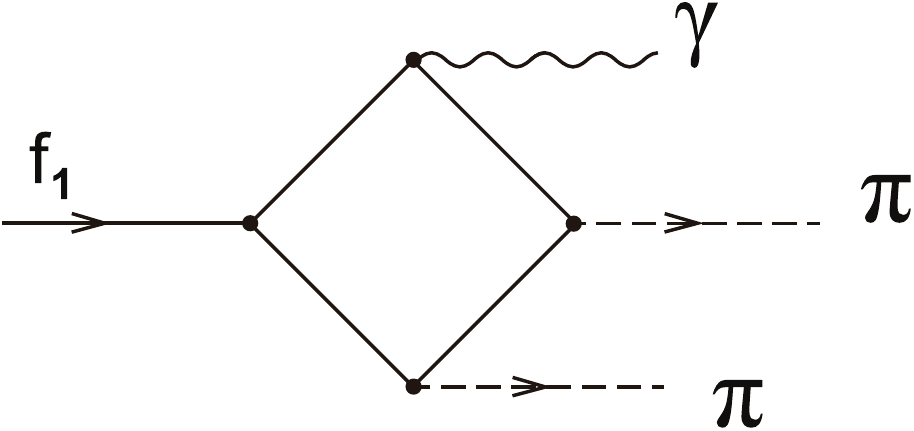}
\caption{The contribution of the contact interaction to the radiative decay amplitude $f_{1}\to\pi^+\pi^-\gamma $.}
\end{figure} 

Let us consider now the amplitudes corresponding to the diagrams plotted on Figs. 2-3. The diagram with the intermediate $a_1$ meson contributes as a contact interaction. The terms with $a_1$ propagators have a factor $p^2$ which is equal zero for real photons. Actually, this is a common feature of the NJL, massive Yang-Mills and hidden local symmetry approaches \cite{Meissner88,Bando88}. In all these models the decay amplitude $a_1\to\pi\gamma$ is zero on the $a_1$ mass shell. The experimental situation is not settled yet (the PDG \cite{Patrignani16} does not give a definite number for the $a_1\to\pi\gamma$ decay width). For this reason we are not going into details here postponing such analyses  for the future, when we will have enough experimental information on this mode. Thus, we get  
\begin{eqnarray}
\label{Ta1}
T^{(a_1)}&=&-i\frac{eg_\rho}{8\pi^2f_\pi^2}e^{\mu\nu\alpha\beta}\epsilon_\beta (l)\epsilon_\alpha^*(p) \nonumber \\ 
&\times&2\kappa m^2\left[1+(1-3a)\kappa m^2\right]l_\mu q_\nu
\end{eqnarray} 
where the 4-vector $q=p_+ - p_-$. Notice that only contributions due to $\vec\pi -\vec a_1$ transitions survived in (\ref{Ta1}). Writing $l_\mu q_\nu$ as a sum $l_\mu q_\nu\to p_\mu q_\nu -2 p_+^\mu p_-^\nu$ one sees that the term $\propto p_+^\mu p_-^\nu$ brakes gauge invariance. Thus there must be other diagrams to restore the symmetry. 

These diagrams are shown in Fig. 3 (we do not show there, but it is assumed, that each pion line represents the direct creation of a pion by the quark-antiquark pair and the indirect one through the $\vec\pi -\vec a_1$ transition). At leading order of the derivative expansion we obtain the amplitude
\begin{eqnarray}
T^{(b)}&=&i\frac{eg_\rho}{8\pi^2 f_\pi^2} e^{\mu\nu\alpha\beta}\epsilon_\beta (l)\epsilon_\alpha^*(p)\left[\frac{ 
p_\mu q_\nu}{Z} \right. \nonumber \\
&-& \left. \kappa m^2 (4-\kappa m^2)p_+^\mu p_-^\nu \right].
\end{eqnarray} 
Now, one can restore the gauge symmetry of the whole amplitude by fixing the parameter $a$. The requirement is to cancel the unwanted $p_+^\mu p_-^\nu$ term of the sum $T^{(a_1)}+T^{(b)}$. It gives $a=5/12$. The rest of the sum will contribute to the amplitude (\ref{amplitude}) in the form
\begin{eqnarray}
&& F_1^{(a_1+b)}=-F_2^{(a_1+b)}=\frac{1}{2}F_3^{(a_1+b)} \nonumber \\
&& =\frac{eg_\rho}{8\pi^2 f_\pi^2}\left[\frac{2-Z}{Z}+\frac{(Z-1)^2)}{8Z^2} \right].
\end{eqnarray}
The numerical value of this contribution is not large $\Gamma^{(a_1 +b)} (f_1\to\pi^+\pi^-\gamma )=5.5\,\mbox{keV}$. Nonetheless, in the next section we will show that the interference of this amplitude with the pure $\rho$ exchange channel is positive and relatively large. It enhances noticeably the final result.  

\section{$f_1(1285)\to\pi^+\pi^-\gamma$ decay width}
\label{DecW}

The rate of the three-body decay $f_1(1285)\to\pi^+\pi^-\gamma$ can be obtained from the standard formula 
\begin{equation}
d\Gamma =\frac{|T|^2}{24m_{f_1}(2\pi )^3}d\varepsilon d\varepsilon_+
\end{equation}
where 
\begin{equation}
|T|^2=\sum_{i\leq j} \mbox{Re} \left(F_i F_j^* \right) T_{ij},
\end{equation}
and
\begin{eqnarray}
T_{11}&=&2m_{f_1}^2 \vec p_+^{\, 2} \nonumber \\
T_{22}&=&2m_{f_1}^2 \vec p_-^{\, 2} \nonumber \\
T_{33}&=&2[(p_+p_-)^2- m_{\pi}^4]+ (\vec p_+ \times\vec p)^2  \nonumber \\
T_{44}&=&- m_{f_1}^4 (\vec p_+ \times\vec p)^2  \nonumber \\
T_{12}&=&4m_{f_1}^2 \vec p_+\vec p_- \nonumber \\
T_{13}&=&4m_{f_1} [m_\pi^2\varepsilon_- - (p_+ p_- )\varepsilon_+ ]  \nonumber \\
T_{23}&=&-4m_{f_1} [m_\pi^2\varepsilon_+ - (p_+ p_- )\varepsilon_- ]  \nonumber \\
T_{14}&=&T_{24}=0 \nonumber \\
T_{34}&=&-2m_{f_1}^2 (\vec p_+ \times\vec p)^2 .
\end{eqnarray}
Notice that 
\begin{eqnarray} 
&&(\vec p_+ \times\vec p)^2=(\vec p_- \times\vec p)^2=(\vec p_+ \times\vec p_-)^2 \nonumber \\
&&=\vec p_+^{\, 2}\vec p^{\, 2}-(\vec p_+\vec p)^2.
\end{eqnarray} 
Here all kinematical variables are given in the rest frame of the $f_1$ meson. With the use of the kinematic invariants $s=(l-p)^2$, $t=(l-p_+)^2$ and $u=(l-p_-)^2$ one can find the boundary of the physical region in the Mandelstam plane. For a given value of $s$ from the closed interval $4m_\pi^2\leqslant s\leqslant m_{f_1}^2$ the boundary is given by the two roots of the quadratic equation. They are
\begin{equation}
t_{\pm}(s)=\frac{1}{2}\left( m_{f_1}^2+2m_\pi^2-s  \pm \sqrt{D(s)} \right),
\end{equation} 
where $m_\pi^2\leqslant t_\pm \leqslant (m_{f_1}-m_\pi )^2$, and
\begin{eqnarray}
D(s)&=&\left(m_{f_1}^2+2m_\pi^2-s\right)^2 \nonumber \\
&-&4m_\pi^2\left(\frac{m_{f_1}^4}{s}-m_{f_1}^2+m_\pi^2\right).
\end{eqnarray}

Collecting above formulae we come to the following result
\begin{equation}
\label{G}
\Gamma = \frac{1}{24m_{f_1}(2\pi )^3} \int\limits_0^{\varepsilon^{\mbox{\tiny max}}}d\varepsilon\int\limits_{\varepsilon_+^{\mbox{\tiny min}}}^{\varepsilon_+^{\mbox{\tiny max}}} d\varepsilon_+ |T|^2
\end{equation}
where 
\begin{eqnarray}
&&\varepsilon^{\mbox{\tiny max}} =\frac{m_{f_1}^2-4m_\pi^2}{2m_{f_1}}, \nonumber \\
&& \varepsilon_+^{\mbox{\tiny min}}=\frac{1}{2}\left[m_{f_1} - \varepsilon \left(1+\sqrt{\Omega(\varepsilon )}\right)\right],  \nonumber \\
&& \varepsilon_+^{\mbox{\tiny max}}=\frac{1}{2}\left[m_{f_1} - \varepsilon \left(1-\sqrt{\Omega(\varepsilon )}\right)\right], \nonumber \\
&& \Omega(\varepsilon )=1-\frac{4m_\pi^2}{m_{f_1}(m_{f_1}-2\varepsilon )}.
\end{eqnarray}

Integrating over energies $\varepsilon_+$ and $\varepsilon$ in (\ref{G}) we obtain the radiative decay width $\Gamma (f_1\to\pi^+\pi^-\gamma ) = 346\, \mbox{keV}$. This includes the contributions of the $\rho^0$ and $a_1^\pm$ exchanges, contact interaction and all possible interferences among these amplitudes. The vector $\rho^0$ resonance gives the major contribution. To be precise, we have found that $\Gamma_{\mbox{\tiny tot}} =(346,4 = 275.6+5.5+65.3)\,\mbox{keV}$ where in the sum we present the contributions of $\rho^0$ (first term), $a_1$ plus contact interaction (second term), and  interferences (third term).     

From Eq. (\ref{G}) one can also derive the spectral effective mass distribution for the system of two charged pions, i.e. a derivative $d\Gamma /d\sqrt{s}$ as a function of $\sqrt{s}$. We plot this function in Fig. 4.  The curve has a clear peak located at $\sqrt{s}=767\,\mbox{MeV}$. If one would neglect a contact interaction the maximum would shift to the point $\sqrt{s}=768\,\mbox{MeV}$. We conclude that the shift due to a contact interaction is rather small, or, in other words, the AAA and AAAV anomalies do not affect much the position of the extremum.

On the other hand, a new empirical information is hoped to give more insight on the structure of the spectral function. In particular, this can shed some light on the axial-vector decay amplitude $a_1(1260)\to\pi\gamma$. There are experimental data indicating that the decay width of  $a_1(1260)\to\pi\gamma$ is not zero, $\Gamma(a_1^+\to\pi^+\gamma)=640\pm 246\,\mbox{keV}$ \cite{Zielinski84}. This may modify the spectral function of Fig. 4 at energies $\sqrt{s}\sim m_{a_1}$. Such theoretical study is possible, however, without precise empirical data this analyses would be too speculative.   

\begin{figure}[t]
\label{figura4}
\includegraphics[width=0.45\textwidth]{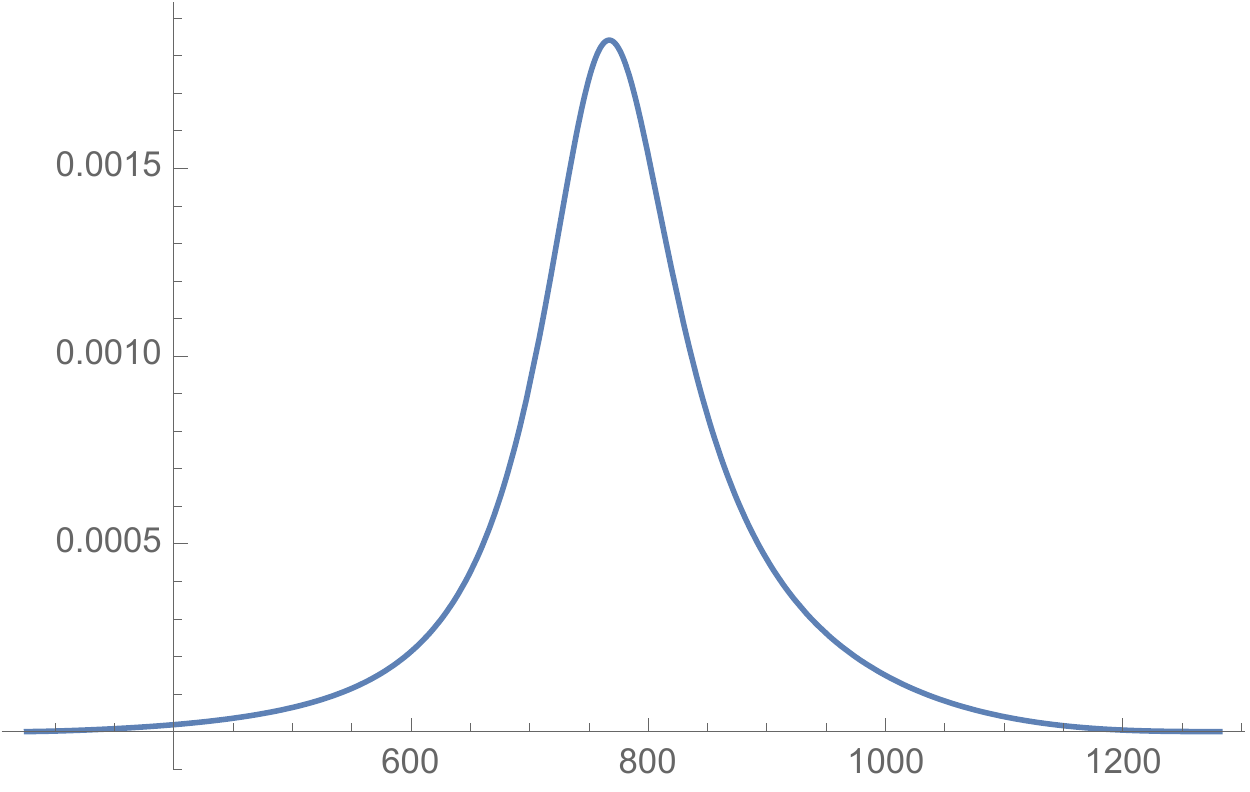}
\caption{The spectral effective mass distribution for the system of two charged pions, $d\Gamma /d\sqrt{s}$, as a function of $\sqrt{s}$ for the radiative decay $f_{1}\to \pi^+\pi^-\gamma$.}
\end{figure} 

\section{Conclusions}
\label{concl}
In this work, we calculate the radiative decay width of the $f_1(1285)$ axial-vector meson into two charged pions. Our main assumptions are that the $f_1(1285)$ state is mostly made of $u$ and $d$ quarks; the decay is governed by the quark triangle AVV anomaly; and the transition $a_1\to\pi\gamma$ is suppressed on the mass shell of the $a_1$ meson. We also neglect the effects of final-state interaction, which can be essential for the $\pi\pi$ system. The latter can be a subject of more refined amplitude analysis, as soon as high statistics empirical data will be available. 

We estimate the contribution of the box  AAAV anomaly and conclude that it is rather small. Nonetheless it affects the result through the positive interference with an intermediate $\rho$ resonance amplitude, enhancing the total value of the decay width on 25\%. 

We also derive the spectral function of the two pion system and show that it has a clear signal of the $\rho (770)$ state. We recommend the experimental study of this spectral function in the future. Our reasoning is that in this way one can extract new important information on the anomaly structure of the amplitude. In particular, to clarify the role of AAA and AAAV anomalies, by studying two pions, and photon spectral functions. It would be also interesting to measure the peak position in the di-pion mass spectrum. A deviation would indicate that only a contribution from the $\rho$ exchange is not sufficient.         

The decay width found, $\Gamma (f_1\to \pi^+\pi^-\gamma) = 346\, \mbox{keV}$, agrees with our previous estimate, $\Gamma_{f_1\to\rho\gamma}=311\ \mbox{keV}$, and is compatible with the recently measured value $\Gamma_{f_1\to\rho^0\gamma}=453\pm 177\ \mbox{keV}$ \cite{Dickson16} within errors. The contributions to $\Gamma (f_1\to \pi^+\pi^-\gamma)$ are quantified as follows: $\Gamma (f_1\to\rho^0\gamma\to \pi^+\pi^-\gamma )=275.6\,\mbox{keV}$, $\Gamma_A (f_1\to\pi^+\pi^- \gamma )=5.5\,\mbox{keV}$, $\Gamma_I (f_1\to \pi^+\pi^-\gamma )=65.3\,\mbox{keV}$, where the subscript $A$ marks the box AAAV anomaly, and $I$ is used for the interference effects. The $\rho$ resonance dominates the amplitude whereas the box anomaly is important through its positive interference with the $\rho$ exchange channel.  

\noindent {\it Acknowledgments:} We are grateful  to S. B. Gerasimov, A. B. Arbuzov, and B. Hiller for interest to our work and useful discussions.

\end{document}